\documentclass[12pt]{article}
\usepackage{epsfig}

\def\beq{\begin{equation}}
\def\eeq#1{\label{#1}\end{equation}}
\def\eeqn{\end{equation}}

\def\beqa{\begin{eqnarray}}
\def\eeqa#1{\label{#1}\end{eqnarray}}
\def\eeqan{\end{eqnarray}}

\let\bar=\overbar

\def\Dslash{\not{\hbox{\kern-4pt $D$}}}
\def\dslash{\not{\hbox{\kern-2pt $\del$}}}

\def\alphas{\alpha_s}
\def\msb{{\bar{\ssstyle M \kern -1pt S}}}

\def\Title#1{\begin{center} {\Large {\bf #1} } \end{center}}

\begin{document}
\noindent DESY 15-241\\
December 2015
\bigskip

\Title{Possible future HERA analyses}

\bigskip\bigskip

%+\addtocontents{toc}{{\it D. Reggiano}}
%+\label{ReggianoStart}

\begin{raggedright}  

{\it Achim Geiser\index{Geiser, A.}\\
DESY, D-22603 Hamburg, Germany\\
E-mail: Achim.Geiser@desy.de\\
\bigskip
Contribution to workshop on Future Physics with HERA data, DESY, Hamburg, November 11-13, 2014.}
\bigskip\bigskip
\end{raggedright}

\begin{abstract}
A variety of possible future analyses of HERA data in the context of the HERA data preservation programme is collected, motivated, and 
commented. The focus is placed on possible future analyses of the existing $ep$ collider data and their physics scope. Comparisons to the original 
scope of the HERA programme are made, and cross references to topics also covered by other participants of the workshop are given.
This includes topics on QCD, proton structure, diffraction, jets, hadronic final states, heavy flavours, electroweak physics, and 
the application of related theory and phenomenology topics like NNLO QCD calculations, low-$x$ related models, nonperturbative 
QCD aspects, and electroweak radiative corrections.  Synergies with other collider programmes are also addressed.
In summary, the range of physics topics which can still be uniquely covered using the existing data is very broad and of considerable 
physics interest, often matching the interest of results from colliders currently in operation. Due to well-established 
data and MC sets, calibrations, and analysis procedures the manpower and expertise needed for a particular analysis is often very much
smaller than that needed for an ongoing experiment. Since centrally funded manpower to carry out such analyses is not available any
longer, this contribution not only targets experienced self-funded experimentalists, but 
also theorists and master-level students who might wish to carry out such an analysis.   
\end{abstract}

\section{Introduction}
\label{sec:intro} 
HERA data taking has ended in 2007, and the primary funding for data analysis has gradually phased out in the period 2012-2014.
Nevertheless, data analysis is continuing in the context of the established DPHEP data preservation programme \cite{DPHEP}. 
From the very beginning the HERA data sets have been unique \cite{Wiik}, and they will remain unique until
a new higher energy electron-positron collider such as LHeC will come into operation. This is expected to happen by the mid 2020s
(after LHC LS3) at the very earliest \cite{LHeC}.
Despite major efforts, many interesting physics topics which can be achieved 
with the HERA data have not been finalised with full statistics, 
or in some cases not even been started. Synergies with LHC results \cite{HERALHC} have already been intensively exploited. 
Further synergy with the 
ongoing LHC programme and with the upcoming lower energy EIC 
programme \cite{EIC} is expected
and will be addressed.

This brief review\footnote{A summary will also be reported elsewhere \cite{PossibleHERA}.} offers a subjective personal collection of such topics, ranging from topics which have arisen since HERA started operations in 1991, all the way up to topics which freshly became of interest recently. Emphasis is placed on their physics motivation and future feasability. 
Hopefully, this will motivate interested people (which do not need to be members of one of the HERA collaborations) to pick up some of 
them and bring them to successful completion within the next decade. This might include theorists 
wishing to compare their predictions to data, experimentalists preparing future programmes who would like to test their analysis ideas on 
real data, and phenomenologists from both theory and experiment which would like to complement results achieved at hadron colliders, 
e.g. at LHC, with results from the cleaner $ep$ initial state.

A description of kinematic variables, structure function definitions and other theory-related terminology as used in this document and an 
explicit comparison of the HERA 
collider and LHC kinematic and detector acceptance ranges can e.g. be found in \cite{PIC08}.
The topics of this contribution will be roughly grouped by the scope of their physics impact, and some of the topics already discussed 
in the Physics at HERA workshop of 1991 \cite{Wiik} will serve as a partial guideline.

\section{Searches for new physics}

Even though a few of the HERA search limits \cite{singletop, Thorne} have not yet been superseded, and a few last ones are currently 
in preparation \cite{Wichmann}, 
due to the limited center-of-mass reach the high energy frontier is clearly not the main scope of the future programme. Also on the intensity
frontier HERA can not compete with dedicated projects. However, in order to find new physics, both the fundamental and many phenomenological parameters of the Standard Model need to be known to the best possible precision. This includes (but is not limited 
to) proton structure, quark masses, coupling constants, and many phenomenological nonperturbative QCD parameters occurring in
fragmentation  and hadronisation. 

The HERA data offer a relatively clean environment to measure such quantities, and are by now 
almost garanteed to be free of biases from new physics. HERA results can thus serve as Standard Candles \cite{Cooper,Glazov} 
against which parameters relevant in new physics 
searches can be calibrated. Several such effects will be addressed among the Standard Model topics in the next sections.  

\section{QCD}
  
The HERA data of $ep$ collisions at $\sqrt{s}=5-320$~GeV still offer 
one of the best opportunities to test and improve our
understanding of the theory of Quantum Chromo Dynamics (QCD).  

QCD can be probed in high-energy electron-proton collisions in several 
different ways. Studies of the structure of the proton yield direct
measurements of the parton density functions at low (HERMES) and high (H1 and ZEUS) energies. 
Collider jet measurements yield 
complementary insights into higher-order QCD corrections, and both taken 
together allow the (experimentally) most precise determinations of the 
strong coupling constant, $\alpha_s$. The study of semi-inclusive final states
such as heavy-flavour production and diffraction yield further complementary 
information about both perturbative and non-perturbative QCD.

\subsection{Proton structure} 

Understanding the structure of the proton in terms of its gluon and quark 
constituents has a threefold physics interest:
\begin{itemize}
\item to understand the properties of the proton in its own right.
\item to understand the details of the underlying theory of QCD.
\item to provide a detailed description of this structure in terms of 
      parton densities, which are an essential input to the physics at 
      hadron colliders such as the Tevatron and the LHC.
\end{itemize}

One of the core goals \cite{Wiik, Engelen} 
and results of HERA, the measurement of inclusive cross sections for Deep Inelastic Scattering (DIS), 
has recently 
been completed \cite{HERAcomb, Cooper}. However, much remains to be done to integrate these data into all global PDF fits, 
fully understand their impact on our knowldege of QCD \cite{Thorne, Moch}, 
and to complement them with measurements of other final states which have not all been completed.  

The topics of the HERMES fixed target spin programme are well covered in separate contributions \cite{HERMES} and are therefore not detailed here. 
Collider topics of future interest include
\begin{itemize}
\item
Completion of  high $x$ studies \cite{Caldwell, Levy}.
\item 
The extraction of improved transverse momentum dependent PDFs \cite{Jung}.
\item
A final extraction of $F_2$ and $F_L$ from the measured \cite{HERAcomb} final reduced cross sections, which depends on theoretical 
input and thus evolves
with improved understanding and precision of QCD theory. A dedicated combined analysis of data allowing an `almost theory independent'
extraction of $F_L$ \cite{FLH1ZEUS} also remains to be done. This will allow improved checks of, and possibly some discrimination 
between, different
theoretical models concerning the treatment of higher order and/or nonperturbative corrections.
\item 
In particular, a more detailed study of the low $x$ part of the data in the context of the resummation of higher order corrections, saturation models, and other low $x$ effects. This interest has been pointed out from the very early days of HERA \cite{Engelen, Bartels1, Bartels2},
 and is still as vibrant as ever \cite{Cooper, Ruspa, Bartels, Motyka}. Questions that can and should be
addressed further include: 

Are we starting to hit the nonperturbative limit?  

Can we make further decisive measurements from existing data?

Can we achieve improved theoretical interpretations from existing results?

Can statements about new physics at high scales be made from the low energy data? \cite{Kowalski}

\end{itemize}

\subsection{Diffraction}

Diffraction is a partially nonperturbative effect which can be particularly well studied at HERA, as was known from very early on
\cite{Ingelman}. To almost everyone's surprise, the diffractive part of the inclusive HERA $ep$ cross sections turned out to be even 
larger than anticipated \cite{Diffraction}, and this phenomenon is not really fully understood up to today, in particular in comparison 
to diffractive cross sections in hadronic collisions. The study of the elastic production of various meson states is a subtopic of great 
interest up to today. Known future topics of interest are
\begin{itemize}
\item
Finalisation of measurements of inclusive diffractive cross sections (most have not been done with full statistics, nor combined), 
and performance of more differential measurements than done so far \cite{Ruspa, Motyka, Volkarova}. 
\item
Finalisation of measurements of the elastic production of $\rho$, $\phi$, $J/\psi$ and other vector mesons \cite{Volkarova}, and comparison to 
improved theory models and to other experiments \cite{McNulty}. 
\item
Measurements of (or limits on) the elastic production of scalar mesons to test the odderon hypothesis. This has been suggested
very early on e.g. for the measurement of $\eta$, $\eta^\prime$ or $\eta_c$ final states \cite{Schaefer}, 
but was never done up to today, even 
though the statistics of the full HERA sample should be sufficient to do so \cite{Sauter}.
\item
The collider DVCS measurements \cite{DVCS} have not yet been updated to the full HERA statistics.
\end{itemize}

\subsection{Jets}

The interest of inclusive and multi-differential jet measurements was again pointed out very early on \cite{Graudenz}, including the 
possibility to use these for measurements of the strong coupling constant $\alpha_s$ \cite{Crombie}. To reduce the dominance of 
the theoretical uncertainty in such measurements, full NNLO QCD calculations are urgently needed. This topic was well covered in 
several contributions to this workshop \cite{Thorne, Volkarova, Gehrmann, Kang}.

A precise knowledge of the strong coupling constant and its evolution with energy scale is one one of the ingredients to test the 
potential unification of coupling constants in so-called grand unification theories. The precision of the knowledge of $\alpha_s$ is 
actually one of the biggest uncertainties in such tests.

Especially on the ZEUS side, inclusive jet measurements have 
not yet been fully completed, and H1 and ZEUS jet measurements have not yet been combined, but have been used successfully
in a combined analysis \cite{HERAcomb}.

Multijet final states \cite{Plaetzer} and Jet related event shapes such as thrust can also be exploited further \cite{Kang}. 

\subsection{Hadronic final states}

Potential measurements of hadronic final states and corresponding perturbative and nonperturbative QCD parameters include 
\begin{itemize}
\item Studies of multi-parton-interactions and other nonperturbative effects \cite{Hautmann}.
\item Studies of the structure of the photon were originally declared a ``central theme of low $Q^2$ studies at HERA'' \cite{Schuler}.
Such measurements have so far been carried out only with 
fractions of the available data sets. The renewed interest in photon structure e.g. in the context of new $e^+e^-$ colliders 
warrants a rich programme of corresponding measurements and theory developments. A direct measurement of gluons in the 
photon could for instance be made from multi-tagged heavy flavour final states \cite{Gladkov}.  
\item QCD instantons are an integral part of standard QCD and therefore must exist if the Standard Model is correct. 
The search for QCD instantons at HERA from high multiplicity final states has so far resulted in inconclusive or negative 
results \cite{Instantons}, but could and should be expanded further \cite{Lohrmann1}.
A new analysis method looking for instanton-induced heavy flavour final states \cite{Bot} has not yet been fully exploited. 
\item Searches for pentaquarks and/or other exotic resonances \cite{Karshon}.
\item The total $\gamma p$ cross section was obviously also a topic of early interest \cite{Lohrmann}. The corresponding 
measurements \cite{gammap} have not yet been fully completed.
\end{itemize}

\subsection{Heavy Flavours}

A recent review of all HERA collider results on open heavy flavour production can be found in \cite{Heavy}. Many of these, like the 
high charm cross section and its potential use to constrain the gluon distribution, were 
anticipated \cite{Ali}, while others, like the measurement of the charm and beauty quark masses \cite{Heavy}, were not. 
Out of these, several topics 
of interest for future measurements remain not fully covered (see also separate contributions \cite{Levy,Thorne}).
\begin{itemize}
\item The question of the existence of intrinsic charm inside the proton, i.e. a charm  contribution not described by perturbatively 
calculable processes, and its potential measurement at HERA, was hotly debated from early on \cite{Ingelman2}. 
At HERA collider energies, there is so far no evidence for intrinsic charm, and upper limits have been set. The corresponding
measurements are mainly limited by the detector tracking acceptance in the forward region. A combined analysis of ZEUS and H1 forward 
muon data using charm semileptonic decays could potentially increase the sensitivity of such analyses.
\item The charm- and beauty-quark masses have already been measured from HERA data \cite{Heavy}, mainly relying on 
deep inelastic scattering final states. Such measurements could be improved further by including the latest available data sets, 
and by completing the NNLO part of the corresponding QCD calculations \cite{Bluemlein}.    
\item Many of the inclusive, differential and multi-differential heavy-flavour cross sections \cite{Heavy} have not yet been measured
 using the full HERA statistics. Some multi-differential and/or multi-tagging measuremnts, which could for instance discriminate better
between quark-quark (boson-gluon-fusion-like) and quark-gluon (compton-like) final states, have not even been started.    
\item Measurements of the charm fragmentation fractions are well advanced \cite{Heavy} and very competitive \cite{Misha}, 
while measurements of charm fragmentation functions have so far only been obtained from HERA I data and have a great potential 
for further improvement. 
\end{itemize}

\section{Electroweak physics}

Although the most important impact of the HERA data occurs in the subfield of QCD, they are obviously also sensitive to electroweak
effects. There are two main lines of interest: Leading order (0 loop) electroweak effects from $W$ and $Z$ exchange which allow the investigation or exploitation of the quark couplings to heavy electroweak bosons. These are particularly interesting in the context of 
the polarised electron data taking during the HERA II running \cite{HERApol, Spiesberger}. And higher order radiative corrections to all processes, including 
QED and electroweak loop corrections.

The issue of radiative corrections was discussed early on \cite{Kramer}, and contributed about 10\% to the 1991 HERA workshop. In practice, QED (or in general electroweak) radiative corrections are often `unfolded' from the measurements such that the cross sections
are reported at QED `Born level'. This is very convenient for QCD predictions, which then do not have to deal with non-QCD higher
order corrections. However, the `unfolding' was often done at Monte Carlo level using leading order corrections only. This has been 
OK for the precision aimed for in past measurements, but will probably not be adequate for future precsison measurements, in 
particular if they are to be integrated with LHC measurements in which NLO electroweak corrections are partially large.
Furthermore, recent advances in including photons into the proton PDF also require a consistent treatment of electroweak effects
at next-to-leading order. In order to allow such corrections to be applied consistently within a given renormalisation scheme, the previously applied leading order corrections must first be undone. This might open a completely new era of PDF fitting, in which documentation 
of exactly which corrections have already been applied to the data sets used, and/or undoing such corrections, might be of major importance.

Topics of future interest include the following.
\begin{itemize}
\item The quark weak couplings have been measured from earlier unpolarised data \cite{weak}, but the corresponding measurements using the HERA II polarised data still need to be completed \cite{Wichmann}.
\item Including the photon structure of the proton, or more generally including NLO electroweak effects into PDF fits to match NNLO QCD
 precision, might need a change of paradigm concerning the treatment of  radiative corrections, as outlined above.
\item The measurement of higher order QED corrections from double photon exchange, e.g. the difference of the Bethe-Heitler dimuon
cross section between $e^+p$ and $e^-p$ collisions, could potentially yield insights into QED/QCD interference of similar interest
as the measurements of the two-photon exchange process in $ep$ elastic scattering by the Olympus collaboration \cite{Olympus}.
\item The study of prompt photon production in $ep$ collisions has recently become a very active field \cite{Bussey} with many future opportunities.
\item Measurements of strangeness in the proton from measurements of charm production in charged  current events are 
within experimental reach of the full HERA data sets, in particular using the secondary vertexing technique,
but have not yet been completed.
\end{itemize}

\section{New theory developments}

New theory developments might trigger the need for a reanalysis of the data. Upcoming full NNLO QCD 
calculations \cite{Gehrmann, Bluemlein}, can, in a first instance, probably be compared to existing data sets. 
Full NLO electroweak radiative corrections, which might be needed to match the QCD NNLO precision, might need a
data reanalysis if the already applied leading order unfolding to Born level, which is part of many `measurements', can not be 
undone reliably (see previous section). The same might be true for analyses trying to exploit new PDF sets including
photons in the proton. New theoretical developments on the topic of exotic Standard Model resonances might trigger 
new dedicated searches for particular predicted final states. 

\section{Synergies with other experimental programmes}

A combined analysis of data from HERA and other sources such as the LHC, the Tevatron, LEP, and fixed target experiments
offers great opportunities for further improvements. This is already standard for global PDF fits, and increasingly for dedicated
aspects of particular PDF studies such as the strangeness contribution \cite{strangeness} or the gluon distribution at 
low x \cite{Prosa}. A combined analysis of other HERA and LHCb data is also very promising \cite{McNulty}.
Such measurements will continue to be of interest. But also on other topics like the measurement
of $\alphas$ from jet data and the measurement of the heavy quark masses from heavy quark data, combined measurements from 
several experimental sources will offer synergy effects. 
Such synergy effects are e.g. within the scope of the PROSA collaboration \cite{Prosa}.
Many of the corresponding analyses can be done using existing cross-section measurements, but as outlined in the previous section a combination with
new theory developments might require a partial data reanalysis.
As an example from the past (which can unfortunately not be done), the possibility of reanalysis of the BCDMS and SLAC fixed target data to 
resolve some apparent discrepancies in PDF fits would have been very helpful.  

In addition, there could also be synergy with future experimental $ep$ programmes: Physicsts and students working e.g. on 
future measurements at EIC \cite{EIC} or LHeC \cite{LHeC} might want to gain experience through related measurements with existing data.

\section{Conclusions}

\bigskip
HERA is currently still one of the best QCD laboratories, and also has a scope for important electroweak measurements. 
Measurements at HERA (at the ``HERA-scale'') are generally in good agreement 
with predictions from Standard Model calculations, and provide, in addition to valuable results in their own right, relevant  
information for measurements at the LHC (the Terascale), such as precise
parametrisations of the parton density functions, precise determinations 
of the strong coupling constant, and insights into the treatment of QCD at 
low $x$, the treatment of quark masses, and the treatment of diffractive
processes. Progress with electroweak effects, in particular with radiative corrections, is also still to be made.
In many cases, some of which have been collected above, the most precise final results with the full HERA 
statistics are still to come. The participation also of non-members of the HERA collaborations to achieve these results and the 
use of synergies with other experimental and phenomenological projects is 
encouraged.

\end{document}